\RequirePackage{fix-cm}
\documentclass{article} 
\usepackage{graphicx}
\usepackage{hyperref}
\usepackage{multirow}
\usepackage{comment}
\usepackage{booktabs}
\usepackage{authblk}
\begin{document}

\title{Quantification of pulmonary involvement in COVID-19 pneumonia by means of a cascade of two U-nets: training and assessment on multiple datasets using different annotation criteria
}
\author{Francesca Lizzi~$^{1,2}$, Abramo Agosti~$^{6}$, Francesca Brero~$^{4,5}$,  Raffaella Fiamma Cabini~$^{4,6}$, Maria Evelina Fantacci~$^{2,3}$, Silvia Figini~$^{4, 11}$, Alessandro Lascialfari~$^{4,5}$, Francesco Laruina~$^{1,2}$, Piernicola Oliva~$^{8,9}$, Stefano Piffer $^{7,10}$ , Ian Postuma~$^{4}$, Lisa Rinaldi~$^{4,5}$, Cinzia Talamonti~$^{7,10}$, Alessandra Retico~$^{2}$\\
\small
$^{1}$ Scuola Normale Superiore, Pisa;\\
$^{2}$ National Institute of Nuclear Physics (INFN), Pisa division, Pisa, IT\\
$^{3}$ Department of Physics, University of Pisa, Pisa, IT;\\ 
$^{4}$ INFN, Pavia division, Pavia, IT;\\
$^{5}$ Department of Physics, University of Pavia, Pavia, IT;\\
$^{6}$ Department of Mathematics, University of Pavia, Pavia, IT;\\
$^{7}$ Department of Biomedical Experimental Clinical Science "M. Serio", University of Florence, Florence, IT \\
$^{8}$ Department of Chemistry and Pharmacy,  University of Sassari, Sassari, IT.\\
$^{9}$ INFN, Cagliari division, Cagliari, IT\\
$^{10}$ INFN, Florence  division, Florence, IT\\
$^{11}$ Department of Social and Political Science, University of Pavia, Pavia, IT;\\
}

\date{Received: date / Accepted: date}

\date{\today}

\maketitle

\begin{abstract}
\textbf{Purpose} 
The automatic assignment of a severity score to the CT scans of patients affected by COVID-19 pneumonia could reduce the workload in radiology departments. This study aims at exploiting Artificial intelligence (AI) for the identification, segmentation and quantification of COVID-19 pulmonary lesions. The limited data availability and the annotation quality are relevant factors in training AI-methods. We investigated the effects of using multiple datasets, heterogeneously populated and annotated according to  different criteria.

\textbf{Methods} We developed an automated analysis pipeline, the $LungQuant$ system, based on a cascade of two U-nets. The first one (U-net$_1$) is devoted to the identification of the lung parenchyma, the second one (U-net$_2$) acts on a bounding box enclosing the segmented lungs to identify the areas affected by COVID-19 lesions. Different public datasets were used to train the U-nets and to evaluate their segmentation performances, which have been quantified in terms of the Dice index. The accuracy in predicting the CT-Severity Score (CT-SS) of the $LungQuant$ system has been also evaluated.

\textbf{Results} Both Dice and accuracy showed a dependency on the quality of annotations of the available data samples. On an independent and publicly available benchmark dataset (COVID-19-CT-Seg), the Dice values measured between the masks predicted by $LungQuant$ system and the reference ones were 0.95$\pm$0.01 and 0.66$\pm$0.13 for the segmentation of lungs and COVID-19 lesions, respectively. The accuracy of 90\% in the identification of the CT-SS on this benchmark dataset was achieved.

\textbf{Conclusion} We analysed the impact of using data samples with different annotation criteria in training an AI-based quantification system for pulmonary involvement in COVID-19 pneumonia. In terms of the Dice index, the U-net segmentation quality strongly depends on the quality of the lesion annotations. Nevertheless, the CT-SS can be accurately predicted on independent validation sets, demonstrating the satisfactory generalization ability of the $LungQuant$.  

\end{abstract}

\section{Introduction}

The task of segmenting the abnormalities of the lung parenchyma related to COVID-19 infection is a typical segmentation problem that can be addressed with methods based on DL. CT findings of patients with COVID-19 infection may include bilateral distribution of ground-glass opacifications (GGO), consolidations, crazy-paving patterns, reversed halo sign and vascular enlargement~\cite{Carotti2020}. Due to the extremely heterogeneous appearance of COVID-19 lesions in density, textural pattern, global shape and location in the lung, an analytical approach is definitely hard to code, whereas it is preferable to learn directly from examples. The potential of DL-based segmentation approaches is particularly suited in this case, provided that a sufficient number of annotated examples are available for training the models.

Few fully automated software tools devoted to this task
have been recently proposed~\cite{Lessmann2021,Fang2021,Ma2020b}.  
Lessmann~{\it et al.}~\cite{Lessmann2021} developed  a U-net model for lesion segmentation trained on semi-automatically annotated COVID-19 cases. The output of this system  was then compbined with the lung lobe segmentation algorithm reported in Xie~{\it et al.}~\cite{Xie2020a}. The approach proposed in Fang~{\it et al.}~\cite{Fang2021} implements the automated lung segmentation method provided in the work of Hofmanninger~{\it et al.}~\cite{Hofmanninger2020}, together with a lesion segmentation strategy based on multiscale feature extraction~\cite{Fang2020}.
The specific problem related to the development of fully automated DL-based segmentation strategies with limited annotated data samples has been explicitly tackled by Ma~{\it et al.}~\cite{Ma2020b}. The authors studied how to train and evaluate a DL-based system for lung and COVID-19 lesion segmentation on a poorly populated samples of CT scans. They also made the data collected for their experiment publicly available and tested their algorithm on a public dataset, allowing for a fair comparison with their system.

In this work we present the DL-based fully automated system to segment both lungs and lesions associated with COVID-19 pneumonia, the $LungQuant$ system, which provides the percentage of lung volume compromised by the infection. We extended the study proposed by Ma~{\it et al.}~\cite{Ma2020b} in two main directions: 1) we investigated the impact of the training annotation style on the prediction across different datasets; 
2) we translated the segmentation problem into a CT-SS assessment problem and we evaluated the reliability of the automated CT-SS assignment across different data samples. 
This paper is structured as follows: we list all the publicly accessible data samples we used to develop and validate the $LungQuant$ system; then, we describe the image analysis pipeline we set up and the training and cross-validation strategies we adopted; finally, we show and discuss the quantification performance  obtained either against a voxel-wise ground truth or in terms of the CT severity scores, according to the information available for each data sample.

\section{Material and Methods}

\subsection{Datasets}
Five public available datasets have been used to train and evaluate our segmentation pipeline. Most of them include image annotations, but each annotation has been associated to patients using different criteria, which are described in the Supplementary Materials. In Table~\ref{tab:summary_data}, a summary of available labels for each dataset is reported. 

\begin{table}[t]
  \centering
  \footnotesize
  \caption{A summary of the datasets used in this study. The CT Severity Score (CT-SS) information is not available for all datasets, but it can be computed for data which has both lung masks and ground-glass opacification (GGO) masks.}
  \begin{tabular}{|p{0.2\textwidth}|c|c|c|c|}
  \hline
  Dataset name & Lung mask & GGO mask & CT-SS & N. of cases \\
  \hline
    Plethora~\cite{Kiser2020}               & Yes & No & No & 402\\
  \hline
  Lung CT Segmentation Challenge~\cite{YangJinzhong2017}    & Yes & No & No & 60\\
  \hline
  COVID-19 Challenge~\cite{An2020}        & No & Yes & No & 199\\
  \hline
  MosMed~\cite{Morozov2020}                & No & No & Inferable & 1110\\
  \hline
  MosMed (annotated subsample)            & No & Yes & Inferable &  50\\
  \hline
  MosMed (in-house annotated subsample)   & Yes & No & Inferable & 91\\
  \hline
  COVID-19-CT-Seg~\cite{Ma2020b}          & Yes & Yes & Inferable & 10\\
    \hline
   \end{tabular}
  \label{tab:summary_data}
\end{table}

\subsection{$LungQuant$: the Deep-Learning based quantification analysis pipeline}
In this section, we describe the fully-automated pipeline we set up for the quantification of lung involvement in patients affected by COVID-19 pneumonia. The analysis pipeline, which is refereed in what follows as the $LungQuant$ system, provides in output the percentage P of lung volume affected by COVID-19 lesions and the corresponding CT severity score (CT-SS=1 for P\textless 5\%, CT-SS=2 for 5\% $\leq$ P\textless 25\%, CT-SS=3 for 25\% $\leq$ P\textless 50\%, CT-SS=4 for 50\% $\leq$ P\textless 75\%, CT-SS=5 for P $\geq$ 75\%).

\begin{figure}[ht!]
    \centering
    \includegraphics[width=0.6\columnwidth]{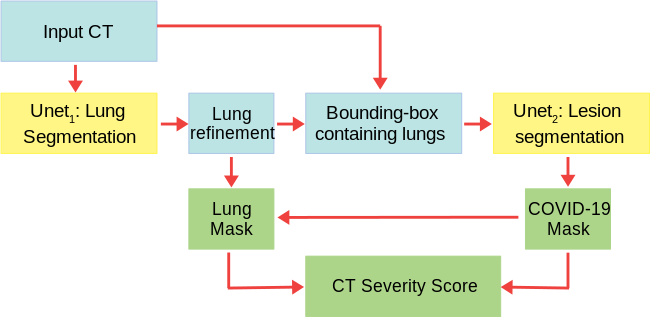}
    \caption{A summary of the whole analysis pipeline: the input CT scans are used to train U-net$_1$, which is devoted to lung segmentation; its output is refined by a morphology-based method. A bounding box containing the segmented lungs is made and applied to all  CT scans for training  U-net$_2$, which is devoted to COVID-19 lesion segmentation. Finally, the output of U-net$_2$ is the definitive COVID-19 lesion mask, whereas  the definitive lung mask is obtained as the union between the outputs of U-net$_1$ and U-net$_2$. The ratio between the COVID-19 lesion mask and the lung mask provides the CT-SS for each patient. }
    \label{fig:pipeline_des}
\end{figure}

A summary of our image analysis pipeline is reported in Fig.~\ref{fig:pipeline_des}.
The central analysis module is a U-net for image segmentation~\cite{Ronneberger2015a} (see sec.~\ref{sec:U-net}), which is implemented in a cascade of two different U-nets: the first network, U-net$_1$, is trained to segment the lung and the second one, U-net$_2$, is trained to segment the COVID lesions in the CT scans. In the following sections, the whole process is described step by step.

\subsubsection{U-net}
\label{sec:U-net}
\begin{figure}[h!]
    \centering
    \includegraphics[width=\columnwidth]{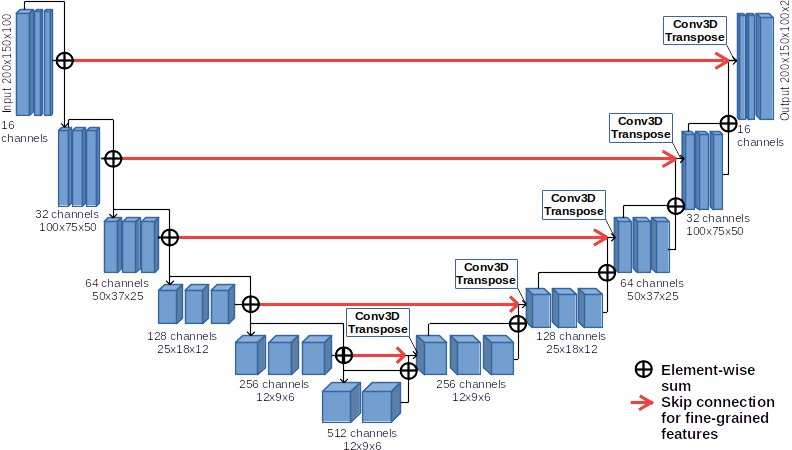}
    \caption{U-net scheme: the neural network is made of 6 levels of depth. In the compression path (left), the input is processed through convolutions, activation layers (ReLu) and instance normalization layers, while in the decompression one (right), in addition to those already mentioned, 3D Transpose Convolution (de-convolution) layers are also introduced. }
    \label{fig:unet}
\end{figure}

For both lung and COVID-19 lesion segmentation, we implemented a fully automated method inspired by the U-net, the fully-convolutional neural networks for image segmentation developed by Ronneberger~{\it et al.}~\cite{Ronneberger2015a}. We implemented a U-net using Keras~\cite{chollet2015keras}, a Python deep-learning API that uses Tensorflow as backend. In Figure~\ref{fig:unet} a simplified scheme of our U-net is reported. 

Each block of layers in the compression path (left) is made by 3 convolutional layers, ReLu activation functions and instance normalization layers. The input of each block is added to the block output in order to implement a residual connection. In the decompression path (right), one convolutional layer has been replaced by a de-convolutional layer to upsample the images to the input size. In the last layer of the U-nets, a softmax is applied to the final feature map and then the loss is computed.

\subsubsection{The U-net cascade for lesion quantification and severity score assignment}
We started by training U-net$_1$, which is devoted to lung segmentation, using the three datasets containing original CT scans and lung masks (see Table.~\ref{tab:summary_data}). The input CT scans, whose number of slices is highly variable, are resampled to matrices of 200x150x100 voxels. The output of U-net$_1$ was then refined using a connected-component labeling strategy, which helps to remove small regions of the segmented mask not connected with the main objects identified as the lungs (see Supplementary Materials for further details). We then built for each CT a bounding box enclosing the morphologically refined segmented lungs, adding a conservative padding of 2.5~cm.
The bounding boxes were used to crop the training images for U-net$_2$, which has the same architecture as U-net$_1$. The cropped images were resized to a matrix of 200x150x100 voxels.
We applied a windowing on the grey-level values of the CT scans to optimize the image contrast for the two segmentation problems we focused on in this analysis. In particular, we selected the [-1000, 1000] HU window range for the U-net$_1$ and  the [-1000, 300] HU range for U-net$_2$. The first window highlights the contrast between the lung parenchyma and the surrounding tissues, whereas the second one enhances the heterogeneous structure of the lung abnormalities related to the COVID-19 infection.
We implemented a data augmentation strategy, relying on the most commonly used data augmentation techniques for DL (see Supplementary Materials for further details) to overcome the problem of having a limited amount of labelled data.

The $LungQuant$ system  returns the infection mask as the output of U-net$_2$ and the lung mask as the union between the output of U-net$_1$ and U-net$_2$. This choice has been made {\it a priori} by design, as U-net$_1$ has been trained to segment the lungs relying on the available annotated data, which are almost totally of patients not affected by COVID-19 pneumonia. Thus, U-net$_1$ is expected to be unable to accurately segment the areas affected by GGO or consolidations; as also these areas are part of the lungs, they should be instead included in the mask. 
Training U-net$_2$ to recognize the COVID-19 lesions on a conservative bounding box containing only the lungs has two main advantages: it allows to restrict the action volume of the U-net to the region where the lung parenchyma (either normal or affected by COVID-19 lesions) is supposed to be, thus avoiding false-positive findings outside the chest; it facilitates the U-net training phase, as the dimensions of the lungs of different patients are normalized, thus the U-net learning process can be focused on the textural patterns characterizing the COVID-19 lesions. 

Finally, once lung and lesion masks have been identified, the $LungQuant$ system computes the percentage of lung volume affected by COVID-19 lesions as the ratio between the total number of voxels of the infection mask and the total number of voxels of the lung mask, and converts it into the corresponding CT severity score.

\subsection{Training details and evaluation strategy for the U-nets}

A detailed description of the metrics used and on the data augmentation strategies implemented is provided in the Supplementary Materials, whereas the data-splitting criterion adopted is described below.

\subsubsection{Cross-validation strategy}
To train, validate and test the performances of each of the two U-nets, we partitioned the available datasets into the training, validation  and test sets, and we evaluated the network performance separately and globally.
U-net$_1$ has been trained and evaluated on CT scans coming from three different datasets: Plethora, MosMed and LCTSC.  U-net$_2$ has been trained and evaluated on samples made of CT scans coming from the COVID-19-Challenge dataset and from the MosMed dataset. 
The amount of CT scan used for train, validation and test sets for each U-net is reported in Table~\ref{tab:CV_sets}. U-net$_2$ has been trained twice, i.e. on both 60\% and 90\% of the CT scans of COVID-19-Challenge and Mosmed datasets to investigate the effect of maximizing training set size on the ability of the system to properly segment the lesions. In the former case, U-net$_2^{60\%}$ training has been evaluated on a validation set made of 20\% of cases and tested on the remaining 20\%. As regard the latter, U-net$_2^{90\%}$, the remaining 10\% of CT scans has been used as validation set. The trained segmentation networks (U-net$_1$ and both U-net$_2^{60\%}$ and U-net$_2^{90\%}$) have been validated on a completely independent validation set consisting of the 10 CT scans of the COVID-19-CT-Seg dataset, which is the only public available dataset containing both lung and infection mask annotations.

\begin{table}[!ht]
\centering
\caption{Number of CT scans assigned to the train, validation (val) and test sets used during the training and performance assessment of   the U-net$_1$ and the U-net$_2$ networks.}
\begin{tabular}{|c|c|c|c|}
\hline
\textbf{U-net$_1$}  &  train & val & test  \\
\hline
Plethora & 319 & 40 & 40 \\
MosMed (91 CT-0) & 55 & 18 & 18 \\
LCTSC & 36 & 12 & 12 \\
\hline
Coronacases &/ &/ & 10 \\
\hline
\hline
\textbf{U-net$_2^{60\%}$}  &  train (60\%) & val (20\%) & test  \\
\hline
COVID-19 Challenge & 119 & 40 & 40 \\
MosMed (50 CT-1) & 30 & 10 & 10 \\
\hline
Coronacases & /&/ & 10 \\
\hline
\hline
\textbf{U-net$_2^{90\%}$}  &  train (90\%)  & val  (10\%) & test \\
\hline
COVID-19 Challenge & 179 & 20 & / \\
MosMed (50 CT-1) & 45 & 5 & / \\
\hline
Coronacases & /& /& 10 \\
\hline
\end{tabular}
\label{tab:CV_sets}
\end{table}

The  $LungQuant$ system has been set up by integrating all analysis modules, as reported in Fig.~\ref{fig:pipeline_des}. In this work we built and analyzed two $LungQuant$ systems, obtained by integrating  alternately U-net$_2^{60\%}$ or U-net$_2^{90\%}$ into the analysis pipeline. The systems have been evaluated in terms of the ability to predict the percentage of affected lung parenchyma and CT-SS on the fully annotated COVID-19-CT-Seg dataset, which is completely independent of the system training phase.

\section{Results}

We report in this section, first, the performance achieved by each of the  segmentation networks we trained, U-net$_1$ and U-net$_2$, then, the quantification performance of the integrated $LungQuant$ system, evaluated on completely independent test sets. We trained both U-nets for 300 epochs on a NVIDIA V100 GPU using ADAM as optimizer, and we kept the  models trained at the epoch where the best evaluation metric on the validation set was obtained. 

\subsection{U-net$_1$: Lung segmentation performance}
\label{sec:Unet1_training}
U-net$_1$ for lung segmentation was trained using three different datasets, as specified in Table~\ref{tab:CV_sets}: the Plethora, a subsample of 91 CT-0 cases of the MosMed dataset and the 60 CT scans of the LCTSC datasets. For the MosMed dataset, as reported in Table~\ref{tab:summary_data}, the lung mask annotations were provided by an in-house developed segmentation software (see Supplementary Materials). Out of the 254 CT scans of the CT-0 MosMed sample, the 91 CT scan we considered here are those on which the in-house segmentation algorithm provided an accurate segmentation, as judged by an experienced medical imaging data analyst.
Then, we tested U-net$_1$ on each of the three independent test sets, and we reported in Table~\ref{tab:lung_results} the performance achieved in terms of Dice values computed between the segmented and the reference masks. The average Dice value obtained on all test samples is also reported. 
We evaluated the lung segmentation performances in three cases: 1) on CT scans and masks resized to the 200x150x100 voxel array size; 2) on CT scans and masks in the original size before undergoing the morphological refinement; 3) on CT scans and masks in the original size and after the morphological refinement. 
Even if segmentation refinement has a small effect on Dice score, as shown in Table~\ref{tab:lung_results}, it is a fundamental step to allow the definition of precise bounding boxes enclosing the lungs, and thus to facilitate the U-net$_2$ learning process.

\begin{table}[!ht]
\centering
\footnotesize
\caption{Performances achieved by  U-net$_1$ in lung segmentation on different test sets, evaluated in terms of the Dice metric at three successive stages of the segmentation procedure.}
\begin{tabular}{|c|c|c|c|}
\hline
Test set & Masks of U-net size & Masks before refinement & Masks after refinement \\ 
& (Dice coefficient) & (Dice coefficient) & (Dice coefficient) \\
\hline
Plethora & 0.96 $\pm$ 0.02 & 0.95 $\pm$ 0.02 & 0.95 $\pm$ 0.04\\
MosMed & 0.97 $\pm$ 0.02 & 0.97 $\pm$ 0.02 & 0.97 $\pm$ 0.02\\
LCTSC & 0.96 $\pm$ 0.03 & 0.95 $\pm$ 0.03 & 0.96 $\pm$ 0.01 \\
Coronacases & 0.96 $\pm$ 0.01 & 0.95 $\pm$ 0.01 &	0.95 $\pm$ 0.01 \\
\hline
\end{tabular}
\label{tab:lung_results}
\end{table}

\subsection{U-net$_2$: COVID-19 lesion segmentation performance}
U-net$_2$ for COVID-19 lesion segmentation has been trained and evaluated separately on the COVID-19-Challenge dataset and on the annotated subset of the MosMed dataset, following the train/validation/test partitioning reported in Table~\ref{tab:CV_sets}. The segmentation performances achieved on the test sets are reported in terms of the Dice metric in Table~\ref{tab:gg_results}. As reported in the table, the performances of U-net$_2$ were evaluated also according to a cross-sample validation scheme.

\begin{table}[!ht]
\centering
\footnotesize
\caption{Performances achieved by  U-net$_2$ in COVID-19 lesion  segmentation, evaluated in terms of the Dice metric. The composition of the train and test sets is reported in Table~\ref{tab:CV_sets}.
}
\begin{tabular}{|c|c|c|c|c|}
\hline
U-net & Trained on & Test set & U-net size & Original CT size\\
 & & & (Dice coefficient) & (Dice Coefficient) \\
\hline

 & \tiny COVID-19 Challenge & \tiny COVID-19 challenge & 0.51 $\pm$ 0.24 & 0.51 $\pm$ 0.25 \\ \cline{2-5} 

 & \tiny COVID-19 Challenge & \tiny  MosMed & 0.39 $\pm$ 0.19 & 0.40 $\pm$ 0.19 \\ \cline{2-5} 

U-net$_2^{60\%}$ & \tiny MosMed & \tiny MosMed & 0.54 $\pm$ 0.22 & 0.55 $\pm$ 0.22\\ \cline{2-5} 

 & \tiny MosMed & \tiny COVID-19 challenge & 0.25 $\pm$ 0.23 & 0.25 $\pm$ 0.23\\ \cline{2-5} 

 & \tiny COVID-19 challenge & \tiny COVID-19 challenge & 0.49 $\pm$ 0.21 & 0.50 $\pm$ 0.21 \\
 & \tiny + MosMed & \tiny + MosMed & & \\
\hline
U-net$_2^{90\%}$ & \tiny COVID-19 Challenge & \tiny COVID-19 Challenge & 0.64 $\pm$ 0.23 & 0.65 $\pm$ 0.23 \\
 & \tiny + MosMed & \tiny + MosMed & & \\
\hline
\end{tabular}
\label{tab:gg_results}
\end{table}

As expected, the U-net$_2$ performances are higher when both the training set and independent test sets belong to the same data cohort. By contrast, when a U-net$_2$ is trained on COVID-19-Challenge data and tested on Mosmed (and the other way around) performances significantly decrease. This effect is due to the fact that the two datasets have been collected and annotated with different criteria. We obtained a better result with the U-net$_{2}$ trained on the COVID-19 Challenge dataset and tested on the MosMed test set, since the network has been trained on a larger data sample and hence it has a higher generalization capability.
When using data from both the COVID-19-Challenge and MosMed datasets in the training and test sets, the Dice index on the test set stands on 0.50 $\pm$ 0.21,  which is similar  to the performance obtained in training  the U-net$_2$  on COVID-19-Challenge data only.
The best segmentation performances have been obtained by the U-net$_2$ trained using the 90$\%$ of the available data, U-net$_2^{90\%}$, which reaches a Dice value of  0.65 $\pm$ 0.23 on the test set.
This result suggests the need to train U-net models on the largest possible data samples in order to achieve higher segmentation performance.

\subsection{Evaluation of the quantification performance of the $LungQuant$ system }
\subsubsection{Evaluation of lung and COVID-19 lesion segmentations}
Once the two U-nets have been trained and the whole analysis pipeline has been integrated in the $LungQuant$ system, we tested it on an independent set (COVID-19-CT-Seg dataset) of CT scans in order to quantify the performances of the whole process. 

\begin{table}[!ht]
\centering
\caption{Performances of the $LungQuant$ system on the independent COVID-19-CT-Seg test dataset. The Dice metric computed between the reference lung and lesion masks and those respectively predicted by the $LunQuant$ system are reported.}
\begin{tabular}{|c|c|c|}
\hline
$LungQuant$ system & Lung segmentation & Infection segmentation\\ 
  & (Dice coefficient) & (Dice coefficient)\\ 
\hline
$LungQuant$ with U-net$_2^{60\%}$ & 0.96 $\pm$ 0.01 & 0.62 $\pm$ 0.09 \\
$LungQuant$ with U-net$_2^{90\%}$ & 0.95 $\pm$ 0.01 & 0.66 $\pm$ 0.13 \\
\hline
\end{tabular}
\label{tab:corona_gg}
\end{table}

\begin{figure}[h!]
    \centering
    \includegraphics[width=0.8\columnwidth]{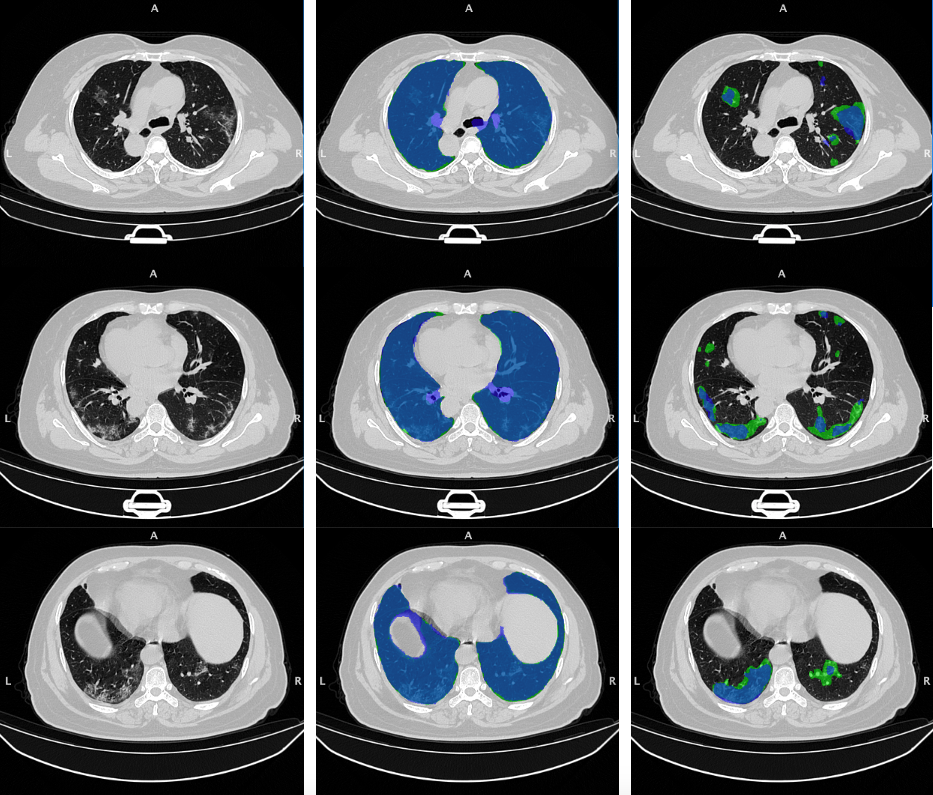}
    \caption{On the rows: three axial slices of the first CT scan on the  COVID-19-CT-Seg test dataset ($coronacases001.nii$) are shown. On the columns: original images (left); overlays between the predicted (blue) and the reference (green) lung (center) and COVID-19 lesion (right) masks. The predicted masks were obtained by the  $LungQuant$ system integrating   U-net$_2^{90\%}$.}
    \label{fig:LungQuant_vs_GS}
\end{figure}

Figure~\ref{fig:LungQuant_vs_GS} allows a visual comparison between the lung and lesion masks provided by the $LungQuant$ systems integrating U-net$_2^{90\%}$ and the reference ones. Three axial slices of the first CT scan of the  COVID-19-CT-Seg test dataset ($coronacases001.nii$) are shown, together with two overlays of the lung and lesion masks, respectively. A very good overlap between the predicted and reference lung masks is observable, whereas a partial overlap occurs between the predicted and reference lesion masks. 

\subsubsection{Percentage of affected lung volume and CT-SS estimation}
\begin{figure}[ht!]
\centering
    {\includegraphics[width=5.5cm]{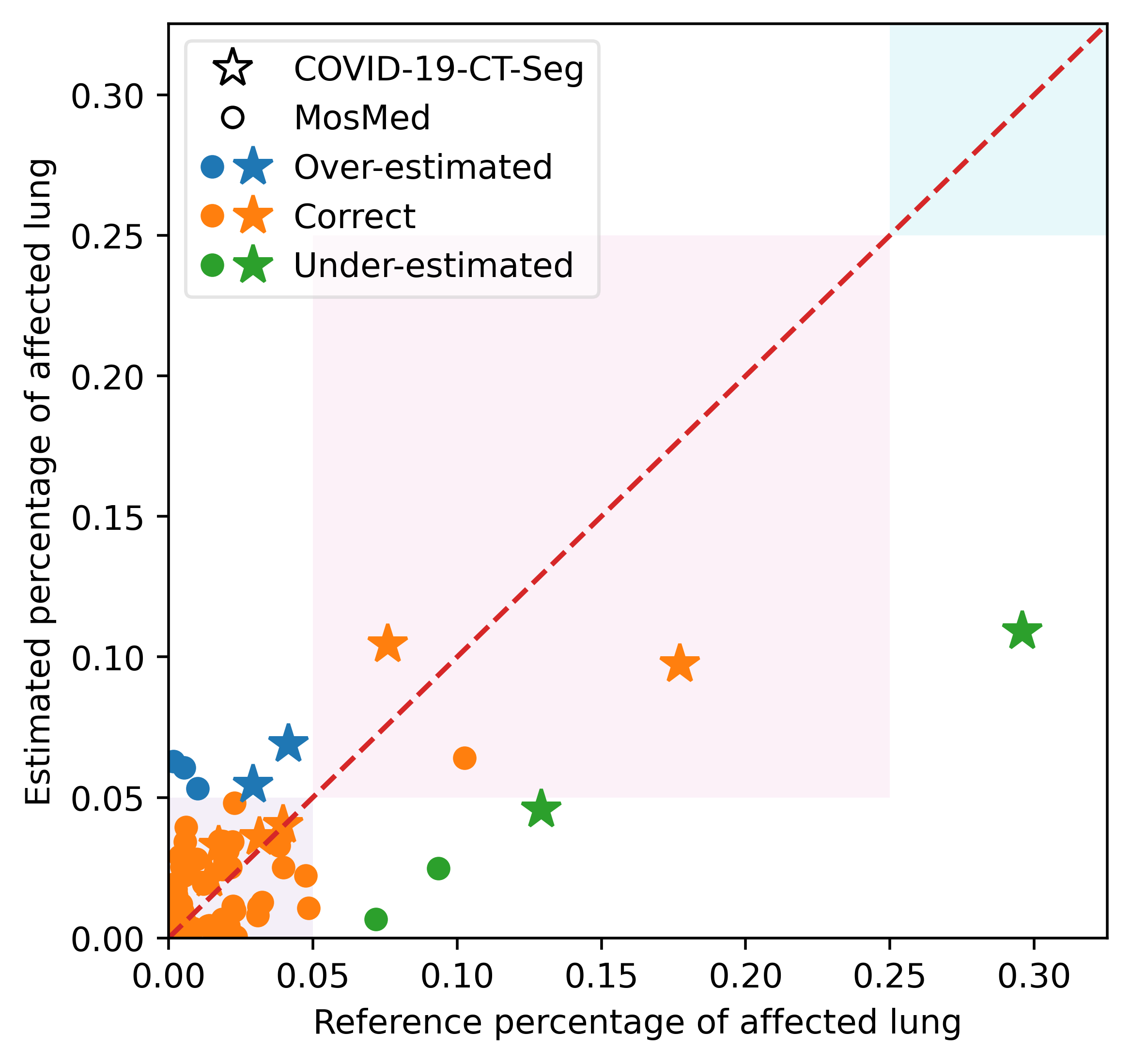}
    \label{fig:sub1}} 
    \hspace{5mm} 
    {\includegraphics[width=5.5cm]{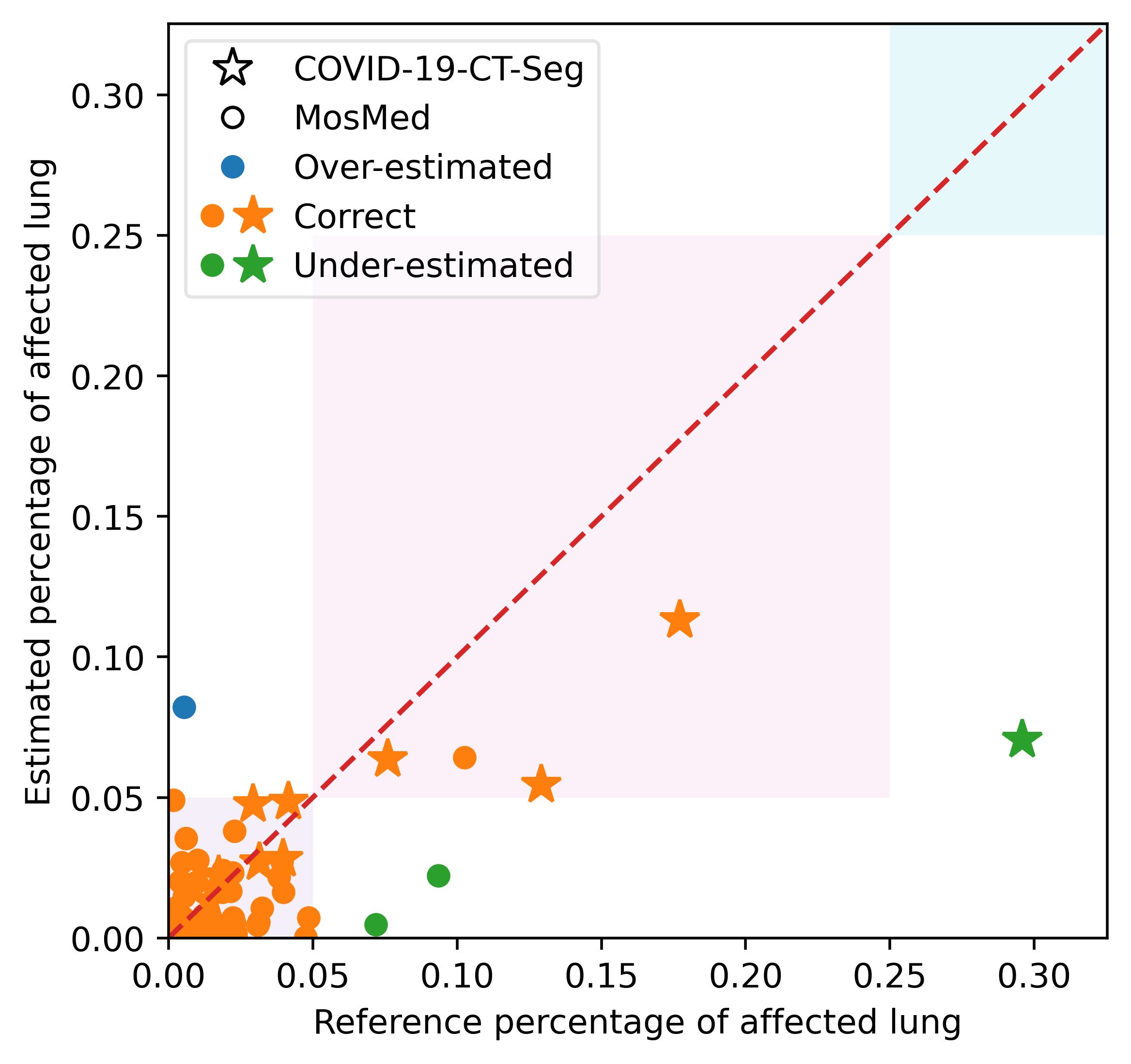}
    \label{fig:sub1}} 
    \caption{Estimated percentages P of affected lung volume versus the ground truth percentages, as obtained by the $LungQuant$ system integrating  U-net$_2^{60\%}$ (left) and U-net$_2^{90\%}$ (right). The colored areas in the plot backgrounds guide the eye to recognize the CT-SS values assigned to each value of P (from left to right: CT-SS=1, CT-SS=2, CT-SS=3. }
    \label{fig:LQ_vs_GS}
\end{figure}

The lung and lesion masks provided by the $LungQuant$ system can be further processed to derive the physical volumes of each mask and the ratios between the lesion and lung volumes. 
We show in Fig.~\ref{fig:LQ_vs_GS} the relationship between the percentage of lung involvement as predicted by the $LungQuant$ system vs. the corresponding values computed on the reference masks, for both the $LungQuant$ systems where the U-net$_2^{60\%}$ and the U-net$_2^{90\%}$ 
were alternatively integrated.
As test samples, we considered the COVID-19-CT-Seg fully independent test dataset, and we complemented it with the partially annotated sample of 50 CT scans of the MosMed collection belonging to the CT-1 class, for which the lesion masks were provided. 
It has to be noticed that this MosMed subsample is not fully independent of the training process since part of this data was used to train the U-net$_2$ networks. As shown in Fig.~\ref{fig:LQ_vs_GS}, the dataset available for this test consists of CT scans with a low percentage P of affected lung, which in most cases is below then 10\%. 
Despite the limited range of P values to carry out this test, an agreement between the $LungQuant$ system output and the reference values is observed for both the systems where either U-net$_2^{60\%}$ or U-net$_2^{90\%}$  were integrated. In terms of the mean absolute error (MAE) among the estimated and the reference P values, we obtained as an average on the 60 test cases: MAE=2.4\% (4.6\% on COVID-19-CT-Seg and 1.9\% on MosMed) for the LungQuant system with  U-net$_2^{60\%}$ and MAE=2.1\% (4.2\% on COVID-19-CT-Seg and 1.7\% on MosMed) for the system with U-net$_2^{90\%}$.

The percentage of lung volume affected by COVID-19 lesions can also be directly converted into the CT-SS values. The accuracy in assigning the correct CT-SS class is reported  in Table~\ref{tab:ct_ss}, together with the number of misclassified cases, for the 10 cases of the COVID-19-CT-Seg dataset and for the subset of 50 annotated MosMed CT scans. As reported in the table, an accuracy of 85\% is achieved for the $LungQuant$ system with  U-net$_2^{60\%}$ and of 93\% for the $LungQuant$ system with  U-net$_2^{90\%}$. 
In all cases, the system misclassifies the examples (9 cases out of 60 for U-net$_2^{60\%}$ and 4 out of 60 cases for the  U-net$_2^{90\%}$) of 1 class at most.

\begin{table}[!ht]
\centering
\caption{Classification performances of the whole system in predicting CT Severity Score on MosMed and Coronacases datasets.}
\begin{tabular}{|c|c|c|c|c|}
\hline
U-net & Dataset & Accuracy & 1-class & 2-class \\ 
 & & & misclassification & misclassification \\
\hline
U-net$_2^{60\%}$ & MosMed & 45/50 & 5/50 & 0\\
\cline{2-5}
& Coronacases &  6/10 & 4/10 & 0\\
\hline
U-net$_2^{90\%}$ & MosMed & 47/50 & 3/50 & 0 \\
\cline{2-5}
& Coronacases & 9/10 & 1/10 & 0\\
\hline
\end{tabular}
\label{tab:ct_ss}
\end{table}

\section{Discussion and Conclusion}

We developed a fully automated quantification pipeline, the $LungQuant$ system, for the identification and segmentation lungs and pulmonary lesions related to COVID-19 pneumonia in CT scans.
The system returns the COVID-19 related lesions, the lung mask and the ratio between their volumes, which is converted  into a CT Severity Score.

The performance obtained against a voxel-wise segmentation ground truth was evaluated in terms of the Dice index, which provides a measure of the overlap between the predicted and the reference masks. The $LungQuant$ system achieved a Dice index of  0.95 $\pm$ 0.01 in the lung segmentation task and of 0.66 $\pm$ 0.13 in segmenting the COVID-19 related lesions on the fully annotated  publicly available benchmark COVID-19-CT-Seg dataset of 10 CT scans.

Regarding the correct assignment of the CT-SS, the $LungQuant$ system showed an accuracy of 93\%, considering the subjects for which the ground truth information was either directly available or derivable within our analysis pipeline. The $LungQuant$ system misclassified only the 7\% of cases of one CT-SS class at most.  
Despite this result is encouraging, it was obtained on a rather small dataset, constituted by COVID-19-CT-Seg and MosMed CT scans, which involves most subjects with low disease severity, thus, a broader validation of larger data sample with more heterogeneous composition in terms of disease severity is required.

Nonetheless, the $LungQuant$ image analysis system we developed can be a useful support tool to assist clinicians in their workflows during the COVID-19 pandemic.

\section*{acknowledgements}
This work has been carried out within the Artificial Intelligence in Medicine (AIM) project funded by INFN (CSN5, 2019-2021), https://www.pi.infn.it/aim.
We are grateful to the staff of the Data Center of the INFN Division of Pisa.
We thank the CINECA  Italian computing center for making available part of the computing resources used in this paper; in particular,  Dr. Tommaso Boccali (INFN, Pisa) as PI of PRACE Project Access \#2018194658 and a 2021 ISCRA-C grant. Moreover, we thank the EOS cluster of Department of Mathematics "F. Casorati" (Pavia) for computing resources. 

\begin{small}

\section*{Conflict of interest}
The authors declare that they have no conflict of interest.

\section*{Ethical approval and informed consent}
All procedures performed in studies involving human participants were in accordance with the ethical standards of the institutional and/or national research committee and with the 1964 Helsinki Declaration and its later amendments or comparable ethical standards.

Informed consent was obtained from all individual participants included in the study.

\end{small}

\bibliographystyle{spmpsci}      
\bibliography{biblio.bib}

\end{document}


\title{SUPPLEMENTARY MATERIALS\\
Quantification of pulmonary involvement in COVID-19 pneumonia by means of a cascade of two U-nets: training and assessment on multiple datasets using different annotation criteria
}
\author{Francesca Lizzi~$^{1,2}$ et al.\\
\small
$^{1}$ Scuola Normale Superiore, Pisa;\\
$^{2}$ National Institute of Nuclear Physics (INFN), Pisa division, Pisa, IT\\
}

\date{Received: date / Accepted: date}

\date{\today}

\maketitle

\section{Additional descriptions of Materials and Methods }

\subsection{Characteristics of the public datasets used in the study}

\subsubsection{The Plethora dataset}
The PleThora dataset~\cite{Kiser2020} is a chest CT scan collection with thoracic volume and pleural effusion segmentations, delineated on 402 CT studies of the Non-Small Cell Lung Cancer (NSCLC) radiomics dataset, available through the The Cancer Imaging Archive (TCIA) repository~\cite{Clark2013}. This dataset has been made publicly available to facilitate  improvement of the automatic segmentation of lung cavities,  which is typically the initial step in the development of automated or semi-automated algorithms  for chest CT analysis. In fact, automatic lung identification struggles to perform consistently in subjects with lung diseases. The PleThora lung annotations have been produced with a U-net based algorithm trained on chest CT of subjects without cancer, manually corrected by a medical student and revised by a radiation oncologist or a radiologist.

\subsubsection{The 2017 Lung CT Segmentation Challenge dataset}
The Lung CT Segmentation Challenge (LCTSC) dataset consists of CT scans of 60 patients, acquired from 3 different institutions and made publicly available in the context of the 2017 Lung CT Segmentation Challenge~\cite{YangJinzhong2017}. Since the aim of this challenge was to foster the development of auto-segmentation methods for organs at risk in radiotherapy, the lung annotations followed the RTOG~1106 contouring atlas. 

\subsubsection{The 2020 COVID-19 Lung CT Lesion Segmentation Challenge dataset}
The 2020 COVID-19 Lung CT Lesion Segmentation Challenge  dataset (COVID-19 Challenge) is a public dataset consisting of unenhanced chest CT scans of 199 patients with positive RT-PCR for SARS-CoV-2~\cite{An2020}. Each CT is accompanied  with the ground truth annotations for COVID-19 lesions. Data has been provided in NIfTI format by The Multi-national NIH Consortium for CT AI in COVID-19 via the NCI TCIA public website~\cite{Clark2013}. Annotations have been made using a COVID-19 lesion segmentation model provided by NVIDIA, which takes a full CT chest volume and produces pixel-wise segmentations of COVID-19 lesions. These segmentations have been adjusted manually by a board of certified radiologists, in order to give 3D consistency to lesion masks. The dataset annotation was made possible through the joint work of Children's National Hospital, NVIDIA and National Institutes of Health for the COVID-19-20 Lung CT Lesion Segmentation Grand Challenge.

The dataset and the annotations have been made available in the context of a MICCAI-endorsed international challenge (https://covid-segmentation.grand-challenge.org/) which had the aim to compare AI-based approaches to automated segmentation of COVID-19 lung lesions. 

\subsubsection{The MosMed dataset}
MosMed~\cite{Morozov2020} is a COVID-19 chest CT dataset collected by the Research and Practical Clinical Center for Diagnostics and Telemedicine Technologies of the Moscow Health Care Department. It includes CT studies taken from 1110 patients. Each study is represented by one series of images reconstructed into soft tissue mediastinal window. MosMed provides 5 labeled categories, based on the percentage of  lung parenchyma affected by COVID-19 lesions. The 5 categories of lung involvement and their correspondence to the CT-SS scale are described in Table~\ref{tab:mosmed_cat}. The 
first category (CT-0) contains cases with normal lung tissue and no CT-signs of viral pneumonia, whereas the other categories contain GGO (CT-1 and CT-2) and both GGO and regions of consolidation in the higher classes (CT-3 and CT-4).

\begin{table}[!ht]
\centering
\caption{MosMed severity categories defined on the basis of the percentage P of lung volume affected by COVID-19 lesions. The correspondence to the CT-SS scale is reported. }
\begin{tabular}{|c|c|rrl|c|}
\hline
MosMed & N. of cases &\multicolumn{3}{c|}{Percentage P of involved} & Corresponding \\
CT category &  & \multicolumn{3}{c|}{lung parenchyma} & CT-SS \\
\hline
0 & 254 &  & P = & 0 & 0 \\
\hline
1 & 684 &\quad 0 & $<$ P $\leq$ & 25 & 1, 2 \\
\hline
2 & 125 &\quad 25 & $<$ P $\leq$ & 50 & 3 \\
\hline
3 & 45 &\quad 50 & $<$ P $\leq$ & 75 & 4 \\
\hline
4 & 2 &\quad 75 & $<$ P $\leq$ & 100 & 5 \\
\hline
\end{tabular}
\label{tab:mosmed_cat}
\end{table}

A small subset of class CT-1 cases (50 patients) had been annotated by expert radiologists with the support of MedSeg software (2020 Artificial Intelligence AS). The annotations consist of binary masks in which white voxels represent both ground-glass opacifications and consolidations. Both CT scans and annotations were provided in NIfTI format. During the DICOM-to-NIfTI conversion process, only one slice out of ten was preserved and, as a result, MosMed CT scans have a reduced total number of  slices with respect to the other datasets.

\subsubsection{The COVID-19-CT-Seg dataset}
The COVID-19-CT-Seg dataset is a collection of CT scans taken from the Coronacases Initiative and Radiopaedia~\cite{Ma2020b}. It contains 20 CT scans tested positive for COVID-19 infection. This public dataset contains both lung and infection annotations. The ground truth has been made in three steps: first, junior radiologists (1-5 years of experience) delineated lungs and infections annotations, then two radiologists (5-10 years of experience) refined the labels and finally the annotations have been verified and optimized by a senior radiologist (more than 10 years of experience in chest radiology). The annotations have been produced with the ITK-SNAP software. Ten CT images of this dataset were provided in 8-bit depth, therefore, we decided to not use them.

\subsection{Additional training details and evaluation strategy for the U-nets}

\subsubsection{Evaluation metrics}

The segmentation performances for both U-nets have been evaluated with the Dice coefficient, computed between the true mask ($M_{true}$) and the predicted mask ($M_{predict}$), as follows;

\begin{equation}
{\rm Dice}_{metric} = \frac{2\cdot|M_{true} \cap M_{predict}|}{|M_{true}| + |M_{pred}|}
\label{eq:1}
\end{equation}

The loss function used to train the U-net for lung segmentation is the DICE loss, defined as follows
\begin{equation}
{\rm Dice}_{ loss} = 1 - \frac{2\cdot|M_{true} \cap M_{pred}|}{|M_{true}| + |M_{pred}|}
\label{eq:2}
\end{equation}
and computed only on the foreground (white voxels). We used this strategy in order to avoid giving excessive weight to the background (black voxels), since the number of black and white voxels is quite unbalanced in favor of the former.

For U-net$_2$, we used a loss function (L) consisting of the sum of the Dice loss and a weighted cross-entropy (CE), defined as follows:

\begin{equation}
{L} = Dice_{loss} + CE_{weighted}
\label{eq:3}
\end{equation}

\begin{equation}
CE_{weighted} = w(x) \sum_{x \in \Omega} log (M_{true}(x) \cdot M_{pred}(x))
\label{eq:4}
\end{equation}
where $w(x)$ is the weight map which takes into account the frequency of white voxels, $x$ is the current sample and $\Omega$ is the training set. 

Since the background class is larger than the foreground class on the order $10^3$, we computed the weight map $w(x)$ for each ground-truth segmentation to increase the relevance of the underrepresented class, following the approach described in~\cite{phan2020resolving}. The weight map was defined as $w(x)=w_0/f_j$ where $f_j$ is the average number of voxels of the $j^\mathrm{th}$ class over the entire training data set ($j = 0,1$) and $w_0$ is the the average between the frequencies $f_j$.

\subsubsection{Data augmentation}
\begin{figure}[ht!]
    \centering
    \includegraphics[width=\columnwidth]{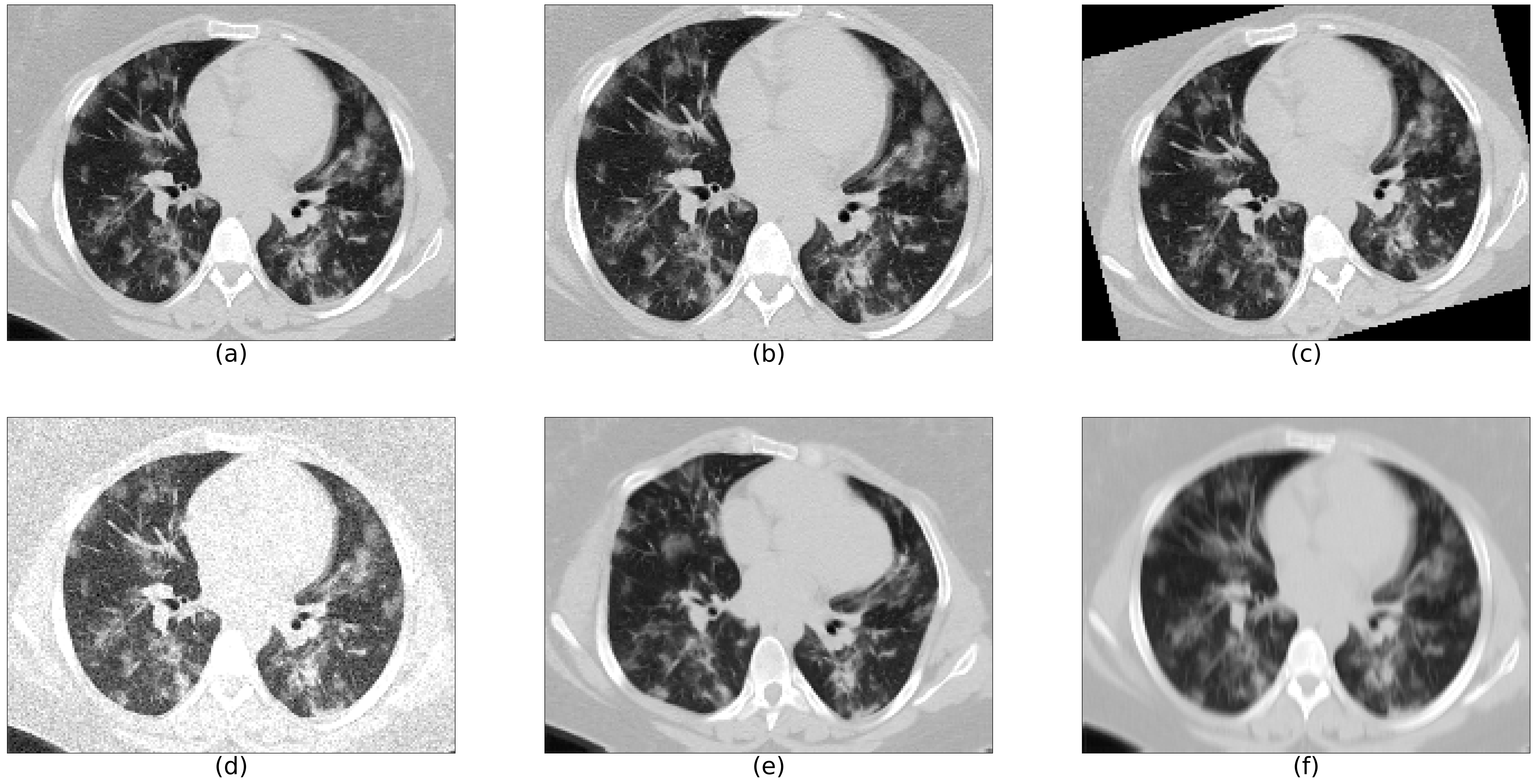}
    \caption{Data augmentation to increase the diversity of dataset: a) Image without data augmentation; b) Zooming; c) Rotation; d) Gaussian noise; e) Elastic deformations; f) Motion blurring.}
    \label{fig:DataAug}
\end{figure}

Data augmentation is a strategy to increase the size of the training set by synthetically generating additional training images through geometric transformations. This technique is particularly important to improve the generalization capability of the model, especially in the case of a limited number of training samples. In our work, we applied data augmentation during the data pre-processing phase (after defining the bounding boxes enclosing the segmented lungs) in order to generate a fixed number of augmented images for each original data. We chose an augmentation factor equal to 2 which means that the number of artificially generated images is twice the number of the original training set.  
For each image in the training set, two of the following geometric transformations were randomly chosen: 
\begin{itemize}
    \item Zooming. The CT image and the ground truth masks were zoomed in the axial plane, using a third-order spline interpolation and the k-nearest neighbor method, respectively. The zooming factor was randomly chosen among the following values: 1.05, 1.1, 1.15, 1.2. \item Rotation. The CT image and the ground truth mask were rotated in the axial plane, using a third-order spline interpolation and the k-nearest neighbor method, respectively. The rotation angle was randomly sampled among the following values: -15°, -10°, -5°, 5°, 10°, 15°. \item Gaussian noise. An array of noise terms randomly drawn from a normal distribution was added to the original CT image. For each image, the mean of the Gaussian distribution was randomly sampled in the [-400, 200] HU range and the standard deviation randomly chosen among 3 values: 25, 50, 75 HU.  
    \item Elastic deformation. An elastic distortion was applied to the original 3D CT and mask arrays following the approach of Simard~{\it et al.}~\cite{Simard2003}. This transformation has two parameters: the elasticity coefficient which we fixed to 12 and the scaling factor, fixed to 1000.  
    \item Motion blurring. Slice by slice, we convolved the CT image with a linear kernel (i.e. ones along the central row and zero elsewhere for a matrix of size $k\times k$) through the function filter2D, defined in the OpenCV Python library~\cite{opencv_library}, keeping the output image size the same as the input image. The filter is applied with a kernel size of 4, 3, and 3, in the anterior-posterior, latero-lateral and cranio-caudal direction, respectively.
\end{itemize}
An example of the application of these augmentation techniques to one CT scan of the dataset is provided in Fig.~\ref{fig:DataAug}.

\subsection{Morphological refinement of U-net$_1$ lung segmentation}
\begin{figure}[h!]
    \centering
    \includegraphics[width=0.8\columnwidth]{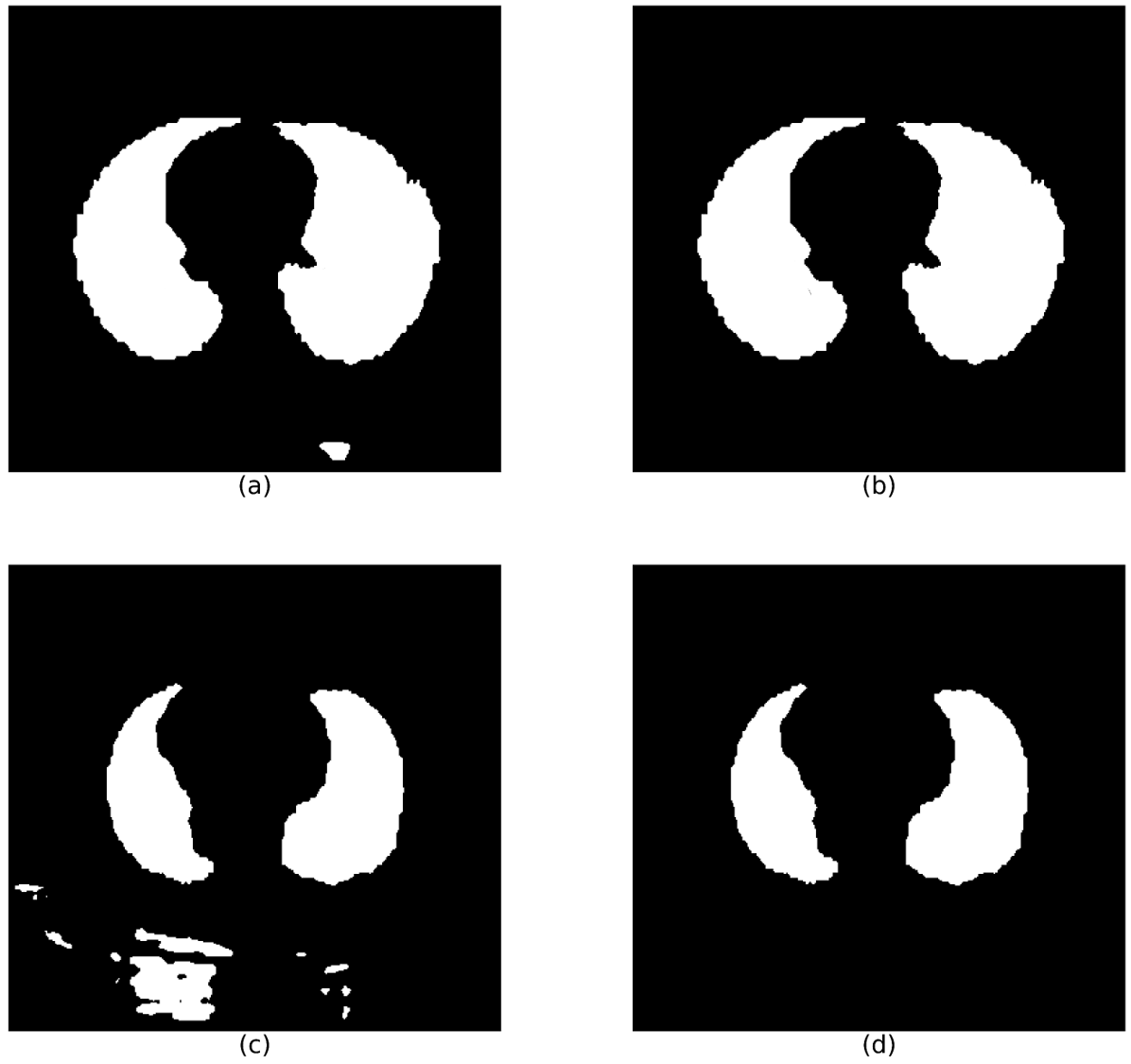}
    \caption{Morphological refinement of the U-net$_1$ output: a) and c) lung masks as generated by  U-net$_1$; b) and d) refined masks after the connected component selection.}
    \label{fig:connected}
\end{figure}

In order to remove false-positive regions (\textit{i.e.} voxels misclassified as lung parts), at first, we identified the connected components in the lung masks generated by U-net$_1$, then, we excluded those components whose number of voxels was below an empirically-fixed threshold. This threshold was set to the 40\% of the foreground mask, and it was reduced to 30\% whether the resulting number of voxels was found to be lower than the 65\% of the initial mask provided by U-net$_1$.
Figure~\ref{fig:connected} shows some examples of how this procedure works on real CT scans. 

\subsection{Generation of a set of reference lung segmentation for model training}
As reported in Table 1 (main paper), the available datasets with lung mask annotations, which were necessary to train the U-net for lung segmentation, are mainly of subjects affected by lung cancer (Plethora and LCTSC datasets). To complement this sample with subjects without lesions, and,  at the same time, to expose to U-net to the acquisition characteristics of the MosMed CT scans, we generated the lung mask annotations for a subset of subjects of the CT-0 MosMed category, i.e. that of subjects without COVID-19 lesions.

An in-house lung segmentation algorithm was developed for this purpose and implemented in $matlab$ (The MathWorks, Inc.). 
It is based on the following steps: 1) CT windowing in the [-1000,1000] HU range; 2) rough segmentation of the lungs on a central coronal slice (Otsu binary thresholding and removal of components connected with the image border) to define the minimum and maximum axial coordinates of the lung region; 3) 2D rough segmentation of the lungs on each axial slice (same procedure as the previous step) to generate a 3D seed mask for the following step; 4) segmentation of the lung parenchyma by an active contour model ($activecontour$ matlab function); 5) filling holes (e.g. vessels and airway walls) with 3D morphological operators ($imclose$ matlab function).

This algorithm, which accurately segments the lung parenchyma in absence of lesions, has very limited performance on CT scans of subjects with COVID-19 lesions.

\bibliographystyle{spmpsci}      
\bibliography{biblio.bib}